\documentclass[aps,prd,showpacs,twocolumn,floatfix]{revtex4}
\usepackage{graphicx}
\usepackage{dcolumn}
\usepackage{bm}
\begin{document}

\def\be{\begin{equation}}
\def\ee{\end{equation}}
\def\bd{\begin{displaymath}}
\def\ed{\end{displaymath}}
\def\ba{\begin{eqnarray}}
\def\ea{\end{eqnarray}}
\def\J{\rm J}
\def\P{\rm P}
\def\C{\rm C}
\def\x{x}
\def\y{y}

\title{\Large
Evidence for a $J/\psi p \bar p$ Pauli Strong Coupling ?}

\author{
T.Barnes,$^{a,b}$\footnote{Email: tbarnes@utk.edu}
X.Li$^{b}$\footnote{Email: xli22@utk.edu}
and
W.Roberts$^{c}$\footnote{Email: wroberts@fsu.edu}\footnote{Notice:
Authored by Jefferson Science Associates, LLC
under U.S. DOE Contract No. DE-AC05-06OR23177. The U.S. Government
retains a non-exclusive, paid-up, irrevocable, world-wide license to
publish or reproduce this manuscript for U.S. Government purposes.}
}
\affiliation{
$^a$Physics Division, Oak Ridge National Laboratory,
Oak Ridge, TN 37831-6373, USA\\
$^b$Department of Physics and Astronomy, University of Tennessee,
Knoxville, TN 37996-1200, USA\\
$^c$Department of Physics and Astronomy, Florida State University,
Tallahassee, FL 32306-4350, USA\\
}

\date{\today}

\begin{abstract}
The couplings of charmonia and charmonium hybrids (generically $\Psi$)
to $p\bar p$ are of great interest in view of future plans to study these
states using an antiproton storage ring at GSI. These low to moderate energy
$\Psi p\bar p$ couplings are not well understood theoretically,
and currently must be determined from experiment.
In this letter we note that the two independent 
Dirac ($\gamma_{\mu}$) and Pauli ($\sigma_{\mu\nu}$) $p\bar p$ couplings 
of the $J/\psi$ and $\psi'$
can be constrained by the angular distribution of
$e^+e^- \to (J/\psi, \psi'\, ) \to p\bar p$ on resonance.
A comparison of our theoretical results to recent unpolarized
data allows estimates of the $p\bar p$ couplings; in the better determined
$J/\psi$ case the data is inconsistent with a pure Dirac ($\gamma_{\mu}$)
coupling, and can be explained by the presence of a $\sigma_{\mu\nu}$ term.
This Pauli coupling may significantly affect the cross section of the PANDA process
$p\bar p \to \pi^0 J/\psi$ near threshold.
There is a phase ambiguity that makes it impossible to uniquely determine 
the magnitudes and relative phase of the Dirac and Pauli couplings from 
the unpolarized angular distributions alone; we show in detail how this can be resolved 
through a study of the polarized reactions.
\end{abstract}


\maketitle

\section{Introduction}
Charmonium is usually studied experimentally through $e^+e^-$ annihilation
or hadronic production, notably in $p\bar p$ annihilation.
The $p\bar p$ annihilation process was employed by the fixed target experiments
E760 and E835 at Fermilab, which despite small production cross sections succeeded in
giving very accurate results for the masses and total widths of the narrow charmonium states
$J/\psi$, $\psi'$, $\chi_1$ and $\chi_2$. A future experimental program of
charmonium and charmonium hybrid production using $p\bar p$ annihilation that is planned by the
PANDA collaboration \cite{PandaTechnicalProgress} at GSI is one of the principal motivations 
for this study.

Obviously the strengths and detailed forms of the couplings of charmonium states
to $p\bar p$ are crucial questions for any experimental program that uses $p\bar p$
annihilation to study charmonium; see for example the predictions for the associated
production processes $p\bar p \to  \pi^0\Psi$ in
Refs.\cite{Gaillard:1982zm,Lundborg:2005am,Barnes:2006ck}.
(We use $\Psi$ to denote a generic charmonium or charmonium hybrid state, and $\psi$
if the state has $J^{PC} = 1^{--}$.)

Unfortunately these low to moderate energy production reactions involve obscure and
presumably rather complicated QCD processes, so for the present they are best
inferred from experiment. In Ref.\cite{Barnes:2006ck} we carried out this exercise
by using the measured $p\bar p$ partial widths to estimate
the coupling constants of the $J/\psi, \psi', \eta_c, \eta_c', \chi_0$ and $\chi_1$
to $p\bar p$, assuming that the simplest Dirac couplings were dominant.
These $\Psi p\bar p$ couplings were then used in a PCAC-like model to give numerical
predictions for several associated charmonium production cross sections of the
type $p\bar p \to \pi^0 \Psi$.

In this paper we generalize these results for the $J/\psi$ and $\psi'$ by relaxing
the assumption of $\gamma_{\mu}$ dominance of the $\psi p\bar p$ vertex.
We assume a $\psi p\bar p$ vertex with both Dirac ($\gamma_{\mu}$) and
Pauli ($\sigma_{\mu\nu}$) couplings, and derive
the differential and total cross sections for
$e^+ e^- \to \psi \to p\bar p$ given this more general vertex.
Both unpolarized and polarized processes are treated.

A comparison of our theoretical unpolarized angular
distributions to recent experimental $J/\psi$ results
allows estimates of both the Dirac and Pauli 
$J/\psi p\bar p$ couplings. There is a phase ambiguity 
that precludes a precise determination of the 
(complex) ratio of the Pauli and Dirac
$J/\psi p\bar p$ couplings from the unpolarized data; 
we shall see that an importance interference effect 
between the Pauli and Dirac terms leads to a strong dependence of the 
unpolarized $p\bar p \to \pi^0 J/\psi$ cross section near threshold 
on the currently unknown phase between these terms.

Determining these couplings is evidently quite
important for PANDA, and can be accomplished through studies of the 
polarized process $e^+ e^- \to J/\psi \to p\bar p$. The 
angular distribution of the unpolarized, self-analyzing process 
$e^+ e^- \to J/\psi \to \Lambda\bar\Lambda$ may also provide complementary 
information regarding the closely related $J/\psi \Lambda\bar\Lambda$ vertex. 
Both of these processes should be accessible at the upgraded BES-III facility.

\section{Unpolarized cross section}
\begin{figure}[h]
\includegraphics[width=0.7\linewidth]{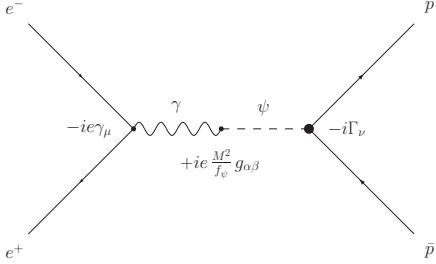}
\caption{The Feynman diagram assumed in this model of
the generic reaction $e^+e^- \to \psi \to p\bar p$.}
\label{fig:diag}
\end{figure}

The Feynman diagram used to model this process is shown in Fig.\ref{fig:diag}.
We assume a vertex for the coupling of a generic $1^{--}$ vector charmonium state
$\psi$ to $p\bar p$ of the form
\begin{equation}
\Gamma_{\mu}^{(\psi p \bar p)} =
g \Big( \gamma_{\mu} +  \frac{i\kappa}{2m}\, \sigma_{\mu\nu} q_{\nu}\Big).
\label{eq:vertex}
\end{equation}
In this paper $m$ and $M$ are the proton and charmonium mass,
$\Gamma$ is the charmonium total width, and
we assume massless initial leptons. Following DIS conventions, $q_{\nu}$
is the four momentum transfer {\it from} the nucleon to
the electron; thus in our reaction $e^+e^-\to p \bar p$ in the c.m. frame,
we have $q = (-\sqrt{s},\vec 0)$.
The couplings $g$ and $\kappa$ are actually momentum dependent form factors,
but since we only access them very close to the kinematic point $q^2 = M^2$ in the reactions
$e^+e^-\to (J/\psi, \psi') \to p \bar p$, we will treat them as constants.

The unpolarized differential and total cross sections for
$e^+e^-\to \psi \to p \bar p$
may be expressed succinctly in terms of the strong $\psi p \bar p$ Sachs form factors
${\cal G}_E = g(1+\kappa s/4m^2)$ and ${\cal G}_M = g(1+\kappa)$.
Both ${\cal G}_E$ and ${\cal G}_M$ are complex above $p\bar p$ threshold, in part because
phases are induced by $p\bar p$ rescattering. If we assume that the lowest-order Feynman diagram 
of Fig.\ref{fig:diag} is dominant, the phase of $g$ itself
is irrelevant, so here we take $g$ to be real and positive. 
$\kappa$ however has a nontrivial phase. 
We express this by introducing a Sachs form factor ratio, with magnitude $\rho \geq 0$ 
and phase $\chi$;
\begin{equation}
{\cal G}_E / {\cal G}_M \equiv \rho e^{i\chi}.
\label{eq:rho_chi_defn}
\end{equation}
The corresponding relation between the Pauli coupling constant $\kappa$ and this Sachs 
form factor ratio is
\begin{equation}
\kappa \equiv |\kappa|\, e^{i\phi_{\kappa}} = 
\frac {\rho e^{i\chi} - 1}{(M^2/4m^2 - \rho e^{i\chi})}
\label{eq:kappa}
\end{equation}
where we have assumed that we are on a narrow resonance, so we
can replace $s$ by $M^2$.

We will first consider the unpolarized process 
$e^+e^-\to \psi \to p \bar p$, and establish what the differential and total 
cross sections imply regarding the $\psi p \bar p$ vertex.
The unpolarized total cross section predicted by Fig.\ref{fig:diag} is 
\begin{equation}
\langle \sigma \rangle =
\frac{4\pi\alpha^2}{3f_{\psi}^2}
\frac{M^4}{s^2}
\frac{(1-4m^2/s)^{1/2}} {[ (s-M^2)^2 + \Gamma^2 M^2 ]}\,
(2m^2 {|\cal G}_E|^2 + s {|\cal G}_M|^2).
\label{eq:sigma}
\end{equation}
(We use angle brackets to denote a polarization averaged quantity.)
Exactly on resonance (at $s=M^2$) this can be
expressed in terms of the $\psi$ partial widths
\begin{equation}
{\hskip -4.7cm}
\Gamma_{\psi \to e^+e^-} = \frac{4\pi \alpha^2 M}{3 f_{\psi}^2}
\label{eq:psi_ee_width}
\end{equation}
and
\begin{equation}
\Gamma_{\psi \to p\bar p } = \frac{(1-4m^2/M^2)^{1/2}}{12\pi M}\, (2m^2 |{\cal G}_E|^2 + M^2 |{\cal G}_M|^2),
\label{eq:psi_pp_width}
\end{equation}
which gives the familiar result
\begin{equation}
\langle \sigma \rangle \big|_{s=M^2} = \frac{12\pi}{M^2} \, B_{e^+e^-} B_{p\bar p }.
\label{eq:sigma_onres}
\end{equation}
Here $B_{e^+e^-}$ and $B_{p\bar p }$ are the
$\psi \to e^+e^-$ and $\psi \to p\bar p$ branching fractions.

Since the (unpolarized) $p\bar p$ width and total cross section on
resonance involve only the
single linear combination $(2m^2 |{\cal G}_E|^2 + M^2 |{\cal G}_M|^2)$,
separating these two strong form factors requires additional information,
such as the angular distribution.
The unpolarized $e^+e^- \to \psi \to p\bar p$ differential cross
section in the c.m. frame is given by
\begin{displaymath}
\langle \frac{d\sigma}{d\Omega}\rangle =
\frac{\alpha^2}{4f_{\psi}^2}
\frac{M^4}{s^2}
\frac{(1-4m^2/s)^{1/2}}
{[ (s-M^2)^2 + \Gamma^2 M^2 ] }
\end{displaymath}
\begin{equation}
\cdot \ \bigg[ 4 m^2 |{\cal G}_E|^2(1-\mu^2) + s |{\cal G}_M|^2 (1+\mu^2)\bigg],
\label{eq:dsigma_dOmega}
\end{equation}
where $\mu = \cos(\theta_{c.m.}).$
This angular distribution is often expressed as
$1+\alpha \mu^2$, where
\begin{equation}
\alpha =
\frac{1-(4m^2/s)\big|{\cal G}_E/{\cal G}_M\big|^2}
     {1+(4m^2/s)\big|{\cal G}_E/{\cal G}_M\big|^2}.
\label{eq:alpha_general_Sachs}
\end{equation}
Inspection of Eqs.(\ref{eq:dsigma_dOmega},\ref{eq:alpha_general_Sachs}) shows that 
one can determine the magnitude $\rho = | {\cal G}_E / {\cal G}_M |$
of the Sachs form factor ratio from the unpolarized
differential cross section, but that the phase $\chi$ of ${\cal G}_E / {\cal G}_M$ 
is unconstrained. 

The undetermined phase $\chi$ implies an unavoidable ambiguity in determining the magnitude and phase 
of the Dirac and Pauli $\psi p \bar p$ couplings $g$ and $\kappa$
from the unpolarized $e^+e^- \to (J/\psi, \psi') \to p\bar p $ angular distribution. We will discuss 
this ambiguity in the next section.

\section{Comparison with Experiment}

\subsection{Summary of the data}

Experimental values of $\alpha$ have been reported by several collaborations. 
The results for the $J/\psi$ are
\begin{equation}
\alpha =
\cases
{
1.45\pm 0.56,  &MarkI~\cite{Peruzzi:1977pb}     \cr
1.7\pm 1.7,    &DASP~\cite{Brandelik:1979hy}      \cr
0.61\pm 0.23,  &MarkII~\cite{Eaton:1983kb}      \cr
0.56\pm 0.14,  &MarkIII~\cite{Brown:1984jq}  \cr
0.62\pm 0.11,  &DM2~\cite{Pallin:1987py}        \cr
0.676\pm 0.036 \pm 0.042,  &BES~\cite{Bai:2004jg}.
}
\label{eq:alpha_Jpsi_expt}
\end{equation}
and for the $\psi'$
\begin{equation}
\hskip -1.1cm
\alpha =
\cases
{
0.67\pm 0.15 \pm 0.04, {\hskip 0.5cm}    &E835~\cite{Ambrogiani:2004uj}     \cr
0.85\pm 0.24 \pm 0.04,                   &BES~\cite{Ablikim:2006aw}.
}
\label{eq:alpha_psip_expt}
\end{equation}
For our comparison with experiment we use the statistically most
accurate measurement for each charmonium state, and combine the errors in quadrature.
This gives experimental estimates for $\alpha$ of $0.676\pm 0.055$ and
$0.67\pm 0.155$ for the $J/\psi$ and $\psi'$ respectively.

\subsection{Testing the pure Dirac hypothesis}

\begin{figure}[b]
\vskip 0.5cm
\includegraphics[width=0.8\linewidth]{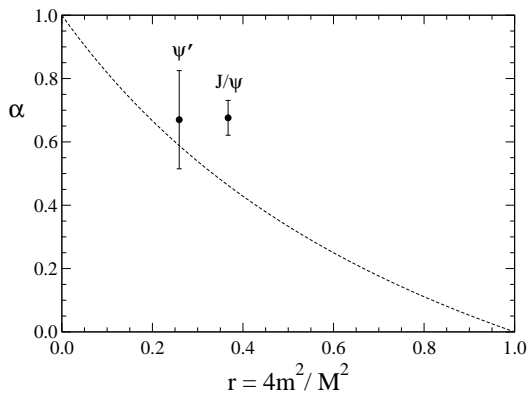}
\caption{The coefficient $\alpha$ observed in the unpolarized
$e^+e^-\to (J/\psi, \psi')\to p\bar p$ angular distributions,
together with the theoretical result $\alpha = (1-r)/(1+r)$ predicted by a pure
Dirac $(\gamma_{\mu})$ $\psi p\bar p$ coupling.}
\label{fig:alpha_nokappa}
\end{figure}

We first examine these experimental numbers using the
``null hypothesis" of no Pauli term, $\kappa = 0$, in which case
$\alpha = (1-r)/(1+r)$, where $r=4m^2/M^2$.
This $\kappa = 0$ formula was previously given by Claudson, Glashow and
Wise~\cite{Claudson:1981fj} and by Carimalo~\cite{Carimalo:1985mw};
the value of $\alpha$ under various theoretical assumptions has been
discussed by these references and by Brodsky and LePage~\cite{Brodsky:1981kj},
who predicted $\alpha=1$.
Fig.\ref{fig:alpha_nokappa} shows these two experimental values
together with the pure Dirac $(\gamma_{\mu})$ formula for $\alpha$.
The $\psi'$ case is evidently consistent with a Dirac
$(\gamma_{\mu})$ $\psi'p\bar p$ coupling at present accuracy,
but the better determined $J/\psi$ angular distribution is
inconsistent with a pure Dirac $J/\psi\, p\bar p$ coupling at the $4\sigma$ level.

The discrepancy evident in Fig.\ref{fig:alpha_nokappa}
may imply the presence of a Pauli term $(\kappa \neq 0)$ in the
$J/\psi\, p \bar p$ vertex. Inspection of our result
for $\alpha$ in the general case (Eq.\ref{eq:alpha_general_Sachs})
shows that one can certainly accommodate this discrepancy by introducing a Pauli term.

\subsection{Determining $\rho = |{\cal G}_E/{\cal G}_M|$ from $\alpha$}

\begin{figure}[h]
\vskip 0.5cm
\includegraphics[width=0.8\linewidth]{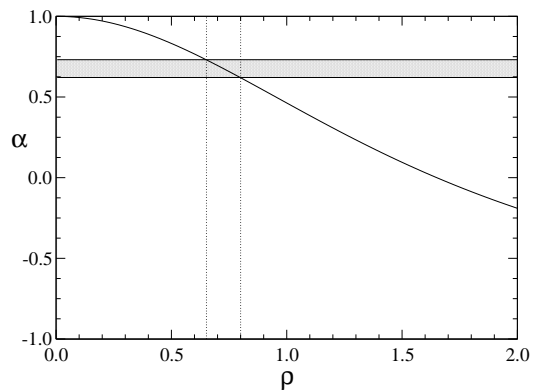}
\caption{The experimental value of the unpolarized $e^+e^-\to J/\psi \to p\bar p$ 
angular coefficient, $\alpha = 0.676\pm 0.055$ (shaded), and the resulting Sachs 
$J/\psi p\bar p$ strong form factor magnitude ratio $\rho = |{\cal G}_E/{\cal G}_M|$ 
(Eq.\ref{eq:rho_solns}).}
\label{fig:alpha_kappa}
\end{figure}

The dependence of the predicted $\alpha$ on $\rho$ at the $J/\psi$ mass
(from Eq.\ref{eq:alpha_general_Sachs})
is shown in Fig.\ref{fig:alpha_kappa}. 
The experimental value $\alpha = 0.676\pm 0.055$ (shown) is consistent with the Sachs form factor
magnitude ratio of
\begin{equation}
\rho = |{\cal G}_E/{\cal G}_M| = 0.726\pm 0.074.
\label{eq:rho_solns}
\end{equation}

In terms of $\rho$ and $\chi$ this completes our discussion: Given the 
unpolarized angular distribution, one obtains a result for 
$\rho = | {\cal G}_E/{\cal G}_M |$ from Eq.\ref{eq:alpha_general_Sachs}, 
but the phase $\chi$ of ${\cal G}_E/{\cal G}_M$ is undetermined. 
However one may ask the more fundamental question of what values of 
the Dirac and Pauli coupling constants $g$ and $\kappa$ in Eq.\ref{eq:vertex} 
are consistent with a given experimental unpolarized angular distribution.  

\subsection{Determining $\kappa$}

\begin{figure}[h]
\vskip 0.5cm
\includegraphics[width=0.8\linewidth]{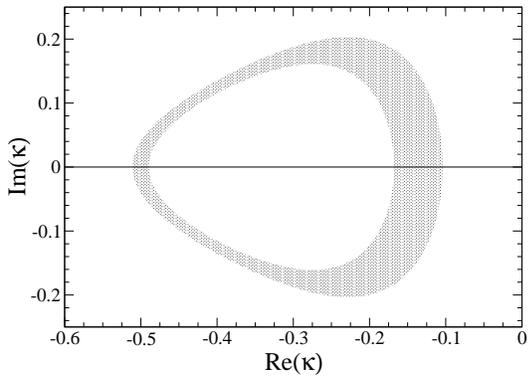}
\caption{The locus of complex $\kappa$ (the $J/\psi p \bar p$ Pauli coupling) 
allowed by the experimental constraint $\rho = 0.726 \pm 0.074$, taken from the 
unpolarized differential cross section for $e^+e^- \to J/\psi \to p \bar p$.} 
\label{fig:blr_fig4}
\end{figure}

First we consider the experimentally allowed values of $\kappa$.
The unpolarized angular distribution provides us with a range of values 
of $\rho$ (Eq.\ref{eq:rho_solns}), but $\chi$ is unconstrained; 
we may combine this information through Eq.\ref{eq:rho_chi_defn} 
to determine the locus of allowed (complex) values of $\kappa$. 
This is shown in Fig.\ref{fig:blr_fig4}. 

For $\chi = 0$, Eq.\ref{eq:rho_chi_defn} implies that $\kappa$ is real 
and negative, and takes on the smallest allowed magnitude. 
As we increase $\chi$ from $0$, the allowed $\kappa$ values proceed 
clockwise, since $\kappa$ initially acquires a 
negative imaginary part. The extreme values of $\kappa$ on the real axis in 
Fig.\ref{fig:blr_fig4} are for $\chi = 0,\pi$, and are 
\begin{equation}
\kappa =
\cases
{
-0.137 \pm 0.032,  & $\chi = 0$    \cr
-0.500 \pm 0.011,  & $\chi = \pi$.
}
\label{eq:kappa_solns}
\end{equation}

\subsection{Determining $g$}

Next we consider the determination of the overall 
$J/\psi \to p \bar p$ vertex strength $g$.
Since the differential and total cross sections for 
$e^+e^- \to J/\psi \to p \bar p$ only involve $g$ through the ratio 
$g/f_{\psi}$,
we must introduce additional experimental data to constrain $g$. 
The partial width for $J/\psi \to p \bar p$ is especially 
convenient in this regard,
since it only involves the strong $J/\psi p \bar p$ vertex, and thus depends 
only on $g$ and $\kappa$ (and kinematic factors). This partial width was
given in terms of the strong Sachs form factors in Eq.\ref{eq:psi_pp_width};
as a function of $g$ and $\kappa$ it is
\begin{equation}
\Gamma_{\psi \to p\bar p} = \frac{1}{3}\frac{g^2}{4\pi} M \sqrt{1-r}\,
\bigg[
1 + \frac{r}{2} + 3\Re(\kappa) + \Big(1+\frac{1}{2r}\Big)|\kappa|^2
\bigg].
\label{eq:psi_ppbar}
\end{equation}

\begin{figure}[h]
\vskip 0.5cm
\includegraphics[width=0.8\linewidth]{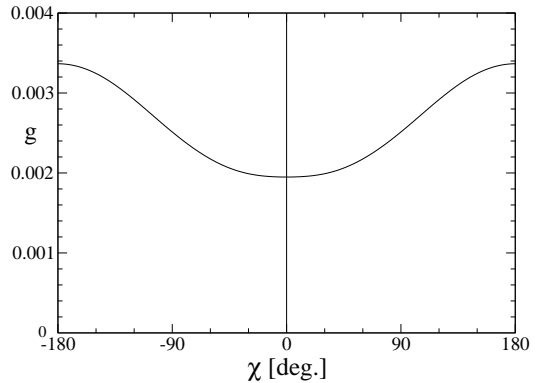}
\caption{The value of the overall $J/\psi p \bar p$ vertex strength 
$g$ implied by the experimental $\Gamma_{J/\psi\to p\bar p}$ and 
$\rho$ as a function of the unknown $J/\psi p \bar p$ Sachs phase $\chi$.}
\label{fig:blr_fig5}
\end{figure}

This generalizes the $\kappa=0$ result given in Eq.27 of 
Ref.\cite{Barnes:2006ck} to a nonzero Pauli coupling.
Using the PDG values \cite{Yao:2006px}
of $\Gamma_{J/\psi} = 93.4 \pm 2.1$~keV
and
$B_{J/\psi\to p\bar p}=(2.17 \pm 0.07)\cdot 10^{-3}$,
Eq.\ref{eq:psi_ppbar} imples a range of values of the overall vertex 
strength $g$ for each value of the (unknown) phase $\chi$. 
This is shown in Fig.\ref{fig:blr_fig5}. There is a range of 
uncertainty in $g$ at each $\chi$ (not shown in the figure), 
due to the experimental errors in $\Gamma_{J/\psi}$, 
$B_{J/\psi\to p\bar p}$ and $\rho$, which is at most $\sim \pm 5 \%$.  

Note that $g$ is bounded by the limits at $\chi = 0$ and $\pi$, 
for which $g \approx 2.0 \cdot 10^{-3}$ and $\approx 3.4 \cdot 10^{-3}$ 
respectively. The allowed values of $g$ are somewhat larger than 
our previous estimate of 
$g = (1.62 \pm 0.03)\cdot 10^{-3}$~\cite{Barnes:2006ck} 
assuming only a Dirac $J/\psi p \bar p$ coupling, as a result of
destructive interference between the Pauli and Dirac terms.

\section{Effect on $\sigma(p\bar p \to \pi^0 J/\psi)$}

The effect of a $J/\psi p\bar p$ Pauli term on the $p\bar p \to \pi^0 J/\psi$ 
cross section may be of considerable interest for the PANDA project, since 
one might use this as a ``calibration" reaction for associated 
charmonium production, and the Pauli term may be numerically important.
Although we have carried out this calculation with the vertex
of Eq.\ref{eq:vertex} for general masses, the full result is rather complicated;
here for illustration we discuss the much simpler massless pion limit.

\begin{figure}[b]
\vskip 0.5cm
\includegraphics[width=0.8\linewidth]{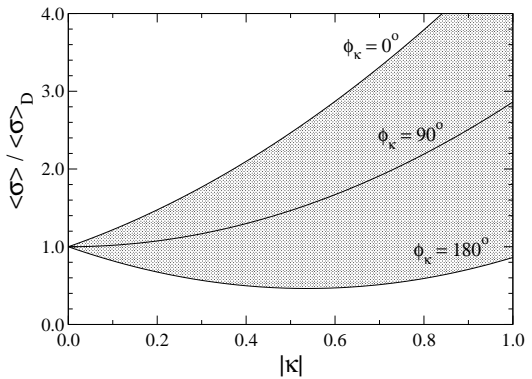}
\caption{The dependence of the unpolarized, near-threshold cross section 
$\langle \sigma (p\bar p \to \pi^0 J/\psi )\rangle$ 
on the (complex) Pauli coupling 
$\kappa = |\kappa| e^{i\phi_{\kappa}}$  
(from Eq.\ref{eq:kappa_effect}).}
\label{fig:sigma_chi}
\end{figure}
For a massless pion the ratio of the unpolarized cross section
$\langle \sigma(p\bar p \to \pi^0 J/\psi) \rangle$
with a Pauli term to the pure Dirac result 
($\gamma_{\mu}$ only, denoted by $D$) is

\begin{displaymath}
\frac{\langle \sigma(p\bar p \to \pi^0 J/\psi )\rangle_{\phantom{D}}}
{\langle \sigma(p\bar p \to \pi^0 J/\psi )\rangle_D}\bigg|_{m_{\pi}=0}
\!\!\! = \bigg[1 + 2\Re(\kappa) + 
\Big(\frac{1}{2}+\frac{M^2}{8m^2}\Big) |\kappa|^2
\end{displaymath}
\begin{equation}
+\,
\frac{(s-M^2)}{4m^2}
\,
\frac{\beta}{\ln\big[(1+\beta)/(1-\beta)\big]}|\kappa|^2\,\bigg],
\label{eq:kappa_effect}
\end{equation}
where $\beta = \sqrt{1-4m^2/s}$ is the velocity of the annihilating $p$ and
$\bar p$ in the c.m. frame. The limit of this cross section ratio 
at threshold is shown in Fig.\ref{fig:sigma_chi} for a range of 
complex~$\kappa$. 

Evidently there is destructive interference for a $\kappa$ with a dominant
negative real part, as is suggested by the unpolarized data. For the value
$\kappa = -0.50$ (the larger solution in Eq.\ref{eq:kappa_solns}) there is
roughly a factor of two suppression in the cross section over the prediction
for a pure Dirac coupling. The suppression however depends strongly on the 
phase of $\kappa$, and for imaginary $\kappa$ has become a moderate 
enhancement. Thus, the near-threshold cross section for 
$p\bar p \to \pi^0 J/\psi$ is quite sensitive to the strength and phase of the 
Pauli coupling; it will therefore be important for PANDA to have an accurate estimate
of this quantity. In the next section we will show how both the magnitude 
and phase of $\kappa$ can be determined in polarized 
$e^+e^- \to J/\psi \to p\bar p$ scattering, and may be accessible at BES. 
 
\section{Polarization Observables}

The relative phase $\chi$ of the 
$J/\psi p\bar p$ Sachs strong form factors 
${\cal G}_E$ and ${\cal G}_M$ 
may be determined
experimentally through a study
of the polarized process $e^+e^- \to J/\psi \to p\bar p$. 
As each of the external particles in this reaction has two 
possible helicity states, there are 16 helicity amplitudes in total.
All the helicity amplitudes to the final $p\bar p$ helicity states
$|p(\pm)\bar p (\pm)\rangle $ are proportional to ${\cal G}_E$, and
all to $|p(\pm)\bar p (\mp)\rangle $ are proportional to ${\cal G}_M$.
In the unpolarized case these are squared and summed,
which leads to a cross section proportional to a weighted sum of
$|{\cal G}_E|^2$ and $|{\cal G}_M|^2$. As we stressed earlier, this implies that 
the phase $\chi$ of ${\cal G}_E / {\cal G}_M$ is not determined by the unpolarized data.

\begin{figure}[h]
\includegraphics[width=0.8\linewidth]{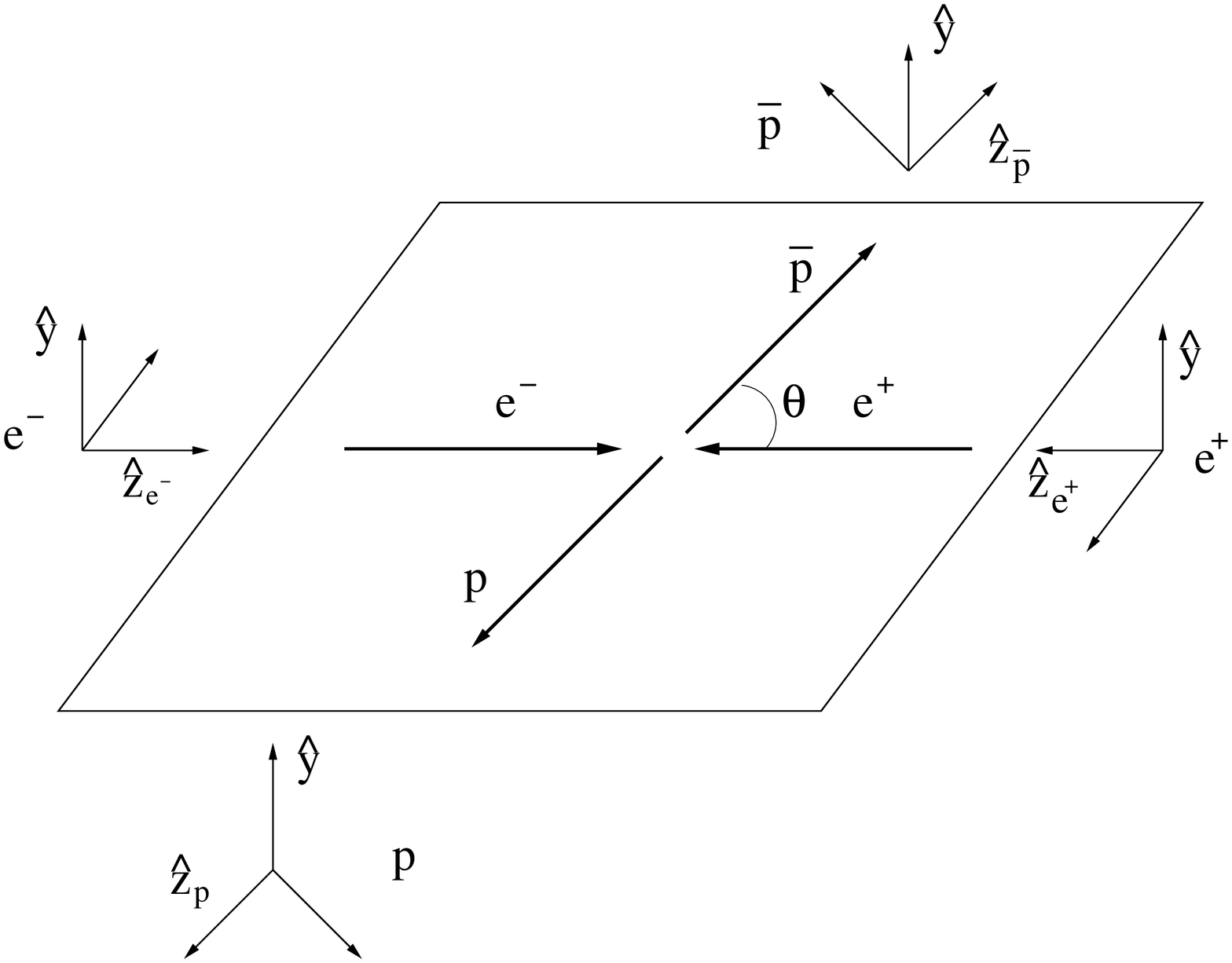}
\caption{Axes used to define the polarization observables.}
\label{fig:axes}
\end{figure}

To show how $\chi$ can be measured in polarized scattering, 
it is useful to introduce the polarization observables discussed by 
Paschke and Quinn~\cite{Paschke:2000mx}. These are angular asymmetries 
that arise when the polarizations of particles are aligned or 
anti-aligned along particular directions. For example, for our reaction
$e^+ e^- \to p\bar p$, $Q(0,0,z,0)$ is the difference of 
two angular distributions, $(d\sigma / d\Omega)_{p\uparrow} - 
(d\sigma / d\Omega)_{p\downarrow}$.
Here we will use $x$ and $y$ for the two transverse axes and 
$z$ for the longitudinal axis (see Fig.\ref{fig:axes}). 
$\hat x$ and $\hat z$ vary with the particle, and
$\hat y$ is chosen to be common to all. An entry of $0$ signifies an 
unpolarized particle. Since there are four possible arguments for 
each particle, $0,x,y$ and $z$, there are $4^4 = 256$ polarization 
observables for this process. Of course there is considerable redundancy, 
since they are all determined by the 16 helicity amplitudes. 
The constraints of parity and charge conjugation reduce this set to 
6 independent helicity amplitudes, and for massless leptons (as we 
assume here) this is further reduced to 3 independent nonzero helicity 
amplitudes.

We introduce the normalized polarization observables 
${\cal Q}_{\epsilon_{e^+}\epsilon_{e^-}\epsilon_{\bar p}\epsilon_p}$
$\equiv $
$Q({\epsilon_{e^+},\epsilon_{e^-},\epsilon_{\bar p}, \epsilon_p})$ /
$Q(0,0,0,0)$,
where $Q(0,0,0,0)$ is the unpolarized differential cross section.
The (nonzero) polarization observables for this process satisfy the relations
\begin{itemize}
\renewcommand{\labelitemi}{$(a)$}
\item
$
\phantom{-} {\cal Q}_{0 0 0 0} =
\phantom{-} {\cal Q}_{x x y y} =
\phantom{-} {\cal Q}_{y y y y} =
-{\cal Q}_{z z 0 0} =
1,
$
\renewcommand{\labelitemi}{$(b)$}
\item
$
\phantom{-} {\cal Q}_{0 0 y 0}=
\phantom{-} {\cal Q}_{x x 0 y}=
\phantom{-} {\cal Q}_{y y 0 y}=
\phantom{-} {\cal Q}_{z z 0 y}=
$
\renewcommand{\labelitemi}{}
\item
$
-{\cal Q}_{0 0 0 y}=
-{\cal Q}_{x x y 0}=
-{\cal Q}_{y y y 0}=
-{\cal Q}_{z z y 0},
$
\renewcommand{\labelitemi}{$(c)$}
\item
$
\phantom{-} {\cal Q}_{x x 0 0} =
\phantom{-} {\cal Q}_{y y 0 0} =
\phantom{-} {\cal Q}_{0 0 y y} =
-{\cal Q}_{z z y y},
$
\renewcommand{\labelitemi}{$(d)$}
\item
$
\phantom{-} {\cal Q}_{z 0 x 0}  =
\phantom{-} {\cal Q}_{z 0 0 x}  =
\phantom{-} {\cal Q}_{y x y z}  =
\phantom{-} {\cal Q}_{y x z y}  =
$
\renewcommand{\labelitemi}{}
\item
$
-{\cal Q}_{0 z x 0} =
-{\cal Q}_{0 z 0 x} =
-{\cal Q}_{x y y z} =
-{\cal Q}_{x y z y},
$
\renewcommand{\labelitemi}{$(e)$}
\item
$
\phantom{-} {\cal Q}_{z 0 z 0}  =
\phantom{-} {\cal Q}_{0 z 0 z}  =
\phantom{-} {\cal Q}_{x y x y}  =
\phantom{-} {\cal Q}_{y x y x}  =
$
\renewcommand{\labelitemi}{}
\item
$
-{\cal Q}_{0 z z 0} =
-{\cal Q}_{z 0 0 z} =
-{\cal Q}_{y x x y} =
-{\cal Q}_{x y y x},
$
\renewcommand{\labelitemi}{$(f)$}
\item
$
\phantom{-} {\cal Q}_{x x x x}  =
\phantom{-} {\cal Q}_{y y x x} =
\phantom{-} {\cal Q}_{z z z z} =
-{\cal Q}_{0 0 z z},
$
\renewcommand{\labelitemi}{$(g)$}
\item
$
\phantom{-} {\cal Q}_{0 0 x x} =
-{\cal Q}_{z z x x}  =
-{\cal Q}_{x x z z}  =
-{\cal Q}_{y y z z},
$
\renewcommand{\labelitemi}{$(h)$}
\item
$
\phantom{-} {\cal Q}_{0 0 x z} =
\phantom{-} {\cal Q}_{x x z x} =
\phantom{-} {\cal Q}_{y y z x} =
\phantom{-} {\cal Q}_{z z z x} =
$
\renewcommand{\labelitemi}{}
\item
$
-{\cal Q}_{x x x z}=
-{\cal Q}_{y y x z}=
-{\cal Q}_{z z x z}=
-{\cal Q}_{0 0 z x},
$
\renewcommand{\labelitemi}{$(i)$}
\item
$
\phantom{-} {\cal Q}_{x y x 0}=
\phantom{-} {\cal Q}_{x y 0 x}=
\phantom{-} {\cal Q}_{z 0 y z}=
\phantom{-} {\cal Q}_{z 0 z y}=
$
\renewcommand{\labelitemi}{}
\item
$
-{\cal Q}_{y x x 0}=
-{\cal Q}_{y x 0 x}=
-{\cal Q}_{0 z y z}=
-{\cal Q}_{0 z z y}.
$
\end{itemize}

\vskip -1.5cm
\begin{equation}
\label{eq:relations}
\end{equation}
Explicit expressions for these observables are given in 
Table~\ref{table:observables}.

\begin{table}[h]
\caption{Nonzero inequivalent polarization observables in $e^+e^-\to J/\psi\to p{\bar p}$.
The function $F$ is $4 - 2(1-\rho^2) \sin^2{\theta}$.}
\begin{center}
\begin{tabular}{c|c}
\hline
Pol. Observable & Result  \\
\hline
${\cal Q}_{0 0 0 0}$ &  $1$ \\
${\cal Q}_{0 0 y 0}$ & $4\rho\,\sin{\chi} \sin{\theta} \cos{\theta}/F$\\
${\cal Q}_{x x 0 0}$ & $2(1-\rho^2)\sin^2{\theta}/F$ \\
${\cal Q}_{z 0 x 0}$ & $4\rho\,\cos{\chi} \sin{\theta}/F$ \\
${\cal Q}_{z 0 z 0}$ & $4\cos{\theta}/F$ \\
${\cal Q}_{0 0 x z}$ & $-4\rho\,\cos{\chi} \sin{\theta} \cos{\theta}/F$ \\
${\cal Q}_{x y x 0}$ & $-4\rho\,\sin{\chi} \sin{\theta}/F$ \\
${\cal Q}_{x x x x}$ & $[4 - 2(1+\rho^2) \sin^2{\theta}]/F$ \\
${\cal Q}_{z z x x}$ & $-2(1 + \rho^2)\sin^2{\theta}/F$ \\
\hline
\end{tabular}
\end{center}
\label{table:observables}
\end{table}

The results in Table~\ref{table:observables} suggest how we may
determine $\chi$ experimentally. 
Inspection of the table shows that only four of the independent 
polarization observables depend on $\chi$; two are proportional to 
$\sin{\chi}$ and two to $\cos{\chi}$. 
Assuming that one knows $\rho$ with sufficient accuracy from the unpolarized data, 
one may then determine $\chi$ unambiguously by extracting $\sin{\chi}$ and $\cos{\chi}$
from the measurement of two of these polarization observables. 

The determination of 
$\sin{\chi}$ is the most straightforward, since it only requires the detection
of a single final polarized particle (for example the proton, through ${\cal Q}_{0 0 y 0}$).
If $\kappa$ is close to real, which corresponds to $\chi \approx 0$ or $\approx \pi$,
this observable may be relatively small. The other polarization observables that are 
proportional to $\sin{\chi}$ involve asymmetries with either one or three particles
polarized; these are given in relations ($b$) and ($i$) of Eq.\ref{eq:relations}.

Determining $\cos{\chi}$ involves measuring double or quadruple polarization 
observables, which are given in relations ($d$) and ($h$) in Eq.\ref{eq:relations}.
In the double polarization case, either one initial and one final polarization 
are measured (such as $e^-$ and $p$) or the polarizations of both final particles 
($p$ and ${\bar p}$) are measured.
In the first case the relevant observables (such as ${\cal Q}_{z 0 x 0}$)
require the initial lepton to have longitudinal ($\pm \hat z$) polarization,
which is difficult to achieve experimentally. In the second case the initial $e^+e^-$ 
beams are unpolarized, and the longitudinal polarization of one final particle and 
the transverse polarization of the other must be measured. Determining the ${\bar p}$ 
polarization may prove to be an experimental challenge.

Of these two general possibilities, the most attractive ``next experiment" beyond 
unpolarized $e^+ e^- \to J/\psi \to p\, \bar p$ scattering may be a 
measurement of the differential cross section with unpolarized leptons and
only the final $p$ polarization detected. This will determine $\sin{\chi}$, 
which specifies $\chi$ up to the usual trigonometric ambiguities. 

Another interesting experimental possibility is to resolve the phase ambiguity 
in unpolarized $e^+ e^- \to J/\psi \to p\, \bar p$ scattering through a study of the 
closely related reaction $e^+ e^- \to J/\psi \to \Lambda \bar \Lambda $,
which has recently been observed by BABAR~\cite{Aubert:2007uf} using the ISR technique.
Since the $\psi p \bar p$ and $\psi \Lambda \bar \Lambda $ couplings are identical in the SU(3)
flavor symmetry limit, a determination of $J/\psi \Lambda \bar \Lambda $ couplings
would suggest plausible $J/\psi p \bar p$ couplings.
This approach has some experimental advantages; as the $\Lambda$ and ${\bar \Lambda}$
decays are self-analyzing, no rescattering of the final baryons is required
to determine their polarization. In addition no beam polarization is required,
since it suffices to measure the (odd-$\rho$) polarization observables ${\cal Q}_{0 0 y 0}$ and 
${\cal Q}_{0 0 x z}$. One may also measure the even-$\rho$ observables
${\cal Q}_{0 0 x x}$ and ${\cal Q}_{0 0 z z}$ as a cross-check of the
result for $\rho$. 

Finally, we note in passing that it may also be possible to measure 
the appropriate polarization observables in the time-reversed reaction 
$p\bar p\to J/\psi \to e^+e^-$.

\section{Summary and Conclusions}

The unpolarized angular distribution for the process 
$e^+ e^- \to J/\psi \to p\, \bar p$, measured recently by the 
BES Collaboration, is inconsistent with theoretical expectations
for a pure Dirac $J/\psi p \bar p$ coupling. In this paper we have 
derived the effect of an additional Pauli-type $J/\psi p\bar p$ coupling, 
and find that this can accommodate the observed angular distribution. 
The $J/\psi p \bar p$ Pauli coupling may significantly affect the cross section
for the charmonium production reaction $ p\bar p \to \pi^0 J/\psi$, which will be
studied at PANDA.
There is an ambiguity in determining the relative Dirac and Pauli
$J/\psi p \bar p$ couplings from the unpolarized
$e^+ e^- \to J/\psi \to p\, \bar p$ data; 
we noted that this ambiguity can be fully resolved through measurements 
of the polarized reaction. The most attractive polarized process to study 
initially appears to be the case of unpolarized initial $e^+e^-$ beams,
with only the final $p$ (transversely) polarized.   
Alternatively, measurement of the required polarization observables may also 
be possible using the time-reversed reaction
$p\bar p \to J/\psi \to e^+e^-$. 
It may also be possible to use self-analyzing processes such as
$e^+ e^- \to J/\psi \to \Lambda \bar \Lambda $ to estimate the Dirac and Pauli
couplings in the closely related $J/\psi \Lambda \bar \Lambda $ vertex.  

\section{Acknowledgements}

We are happy to acknowledge useful communications with
W.M.Bugg, V.Ciancolo, F.E.Close, S.Olsen, J.-M.Richard, K.Seth, S.Spanier, 
E.S.Swanson, U.Wiedner,C.Y.Wong and Q.Zhao
regarding this research. 
We also gratefully acknowledge the support of the Institute of High Energy Physics 
(Beijing) of the Chinese Academy of Sciences, the Department of Physics and 
Astronomy at the University of Tennessee, and the Department of Physics, 
the College of Arts and Sciences, 
and the Office of Research at Florida State University.
This research was supported in part by the U.S. Department of Energy 
under contract DE-AC05-00OR22725 at Oak Ridge National Laboratory.

\vfill\eject

\end{document}